\documentclass[journal,10pt,onecolumn]{IEEEtran} 

\usepackage{cite}
\usepackage{graphicx}
\graphicspath{{figures/}}
\DeclareGraphicsExtensions{.eps}

\usepackage[cmex10]{amsmath}
\usepackage{array}

\usepackage[tight,footnotesize]{subfigure}

\usepackage{url}

\begin{document}
\title{Glow and dust in plasma boundaries}

\author{Victor~Land,~Angela~Douglass,~Ke~Qiao,~Zhuanhao~Zhang,\\~Lorin~Matthews,~Truell~Hyde
\thanks{V.L. is with Foundation for Fundamental Research on Matter, Van Vollenhoven laan 659, 3527 JP, Utrecht, the Netherlands}
\thanks{A.D. is with Ouachita Baptist University, Arkadelphia, AR}
\thanks{Remaining authors are with the Center for Astrophysics, Space Physics and Engineering Research at Baylor University in Waco, TX, 76798 USA, e-mail: truell\_hyde@baylor.edu}
\thanks{Manuscript received June, 2012; revised.}
\thanks{This work is supported by NSF grant \#0847127}}

\markboth{IEEE Trans. Plasma Scie,~Vol.~X, No.~X, Month~2013}%
{Land \MakeLowercase{\textit{et al.}}: Glow and dust in plasma boundaries}

\IEEEpubid{0000--0000/00\$00.00~\copyright~2013 IEEE}

\maketitle
\pagestyle{empty}
\thispagestyle{empty} 
\begin{abstract}
The sheath region is probed in different complex plasma experiments using dust particles in addition to measurement of the optical emission originating from the plasma. The local maximum in optical emission coincides with the breaking of quasi-neutrality at the sheath boundary as indicated by the vertical force profile reconstructed from dust particle trajectories, as well as by the local onset of dust density waves in high density dust clouds suspended in a dielectric box.
\end{abstract}


\section{Introduction}

\IEEEPARstart{P}{lasma} discharges within a confined volume immediately build up boundary layers, called \emph{sheaths}, in front of plasma-facing surfaces. The higher mobility of the electrons as compared to the ions (due to the much smaller mass of the electrons) causes these surfaces to obtain a negative potential. Meanwhile, the interior region of the plasma, called the \emph{bulk}, typically becomes electropositive (when negative ions do not play a role). Therefore, a potential drop proportional to a few times the electron temperature ($T_e$) occurs accross the width of the boundary layer (or \emph{sheath}). The sheath thickness is generally on the order of a few to ten times the electron Debye length, corresponding to lengths of a few millimeters to a centimeter. Thus, the electric field in the sheath typically reaches magnitudes of several thousands of Volts per meter and is directed towards the boundary surfaces.

Macroscopic particles suspended in a plasma collect electrons and ions from the plasma. These particles charge negatively due to the higher mobility of the electrons. The particles can therefore be suspended against the gravitational force by the strong electric field within the sheath. This effect is exploited in complex plasma experiments to levitate micrometer-sized particles, called dust. Dusty plasmas are used as analogue systems to study many fascinating topics from different fields, such as crystallization, phase transitions, waves and even turbulent flows, both in gravity and micro-gravity conditions \cite{Bonitz2010}.

A basic understanding of these systems requires knowledge of the charge-to-mass ratio of the particles, as well as the electric field profile in the sheath. However, a full understanding involves even more effects due to the boundary, including the ion drag force, ion wake effects, and thermal effects. The direct measurement of these variables is difficult, to say the least, due to the small volume involved (making spectroscopic studies very demanding), the perturbative nature of probes, and the fact that the electrostatic force depends on both the charge and the electric field, requiring perturbative methods to decouple them \cite{Melzer1999,Tomme2000}. 

In this paper, we investigate optical emission in the sheath region of low-pressure complex plasma experiments and compare that to the simultaneously measured force profile obtained from particle dropping experiments, which have previously been employed to locate the sheath edge (the breaking of quasi-neutrality) \cite{Douglass2012}. Furthermore, the onset of dust density waves is investigated together with the emission originating from the plasma. 

\section{Experimental results}
The experiments discussed in this paper were performed in two Gaseous Electronics Conference (GEC) reference cells, modified to allow for complex plasma experiments. The cells are capacitively powered at a radio-frequency (RF) of 13.56 MHz. All experiments were performed in argon plasma and the system pressure was fixed at either 20 Pa (first experiment) or 40 Pa (second experiment). The dust particles are spherical melamine-formaldehyde (MF) particles ($\rho$ = 1510 kg m${}^{-3}$). In the first experiment, described in Section \ref{sec:glowandforce}, particles with a diameter of 8.89 $\mu$m were used. In the second experiment, described in Section \ref{sec:glowandddw}, 6.37 $\mu$m diameter particles were suspended within a glass box placed on top of the lower, powered electrode. A detailed characterization of the cell used in the first experiment can be found in \cite{Land2009}, while the cell used in the second experiment is described in \cite{Zhang2010}. Each experiment used a Photron 1024 high-speed camera operating between 500 and 1000 frames per second to image the system. \IEEEpubidadjcol

\subsection{Glow and the force profile}\label{sec:glowandforce}

In a recent study \cite{Douglass2012}, the vertical electric force, $F_E(z) = Q_d(z)E_z(z)$, acting on particles dropped in a discharge was reconstructed by solving the equation of motion of a particle:

\begin{equation}\label{eq:eq_of_motion}
m_d \ddot{z} = F_E(z) - m_dg - m_d\beta\dot{z},
\end{equation}

\noindent where $g$ is the gravitational acceleration and $\beta$ is the Epstein drag coefficient \cite{Epstein1924}. The Epstein drag coefficient is given by

\begin{equation}\label{eq:beta}
\beta=\frac{8}{\pi}\frac{P_{gas}}{{\rho}_dav_{th,gas}},
\end{equation}

\noindent where $P_{gas}$ is the neutral pressure, ${\rho}_d$ is the dust mass density, $a$ is the particle radius, and $v_{th,gas} = \sqrt{k_bT_{gas}/m_{gas}}$.

\begin{figure}[!htbp]
\centering
\includegraphics[width=3.5in]{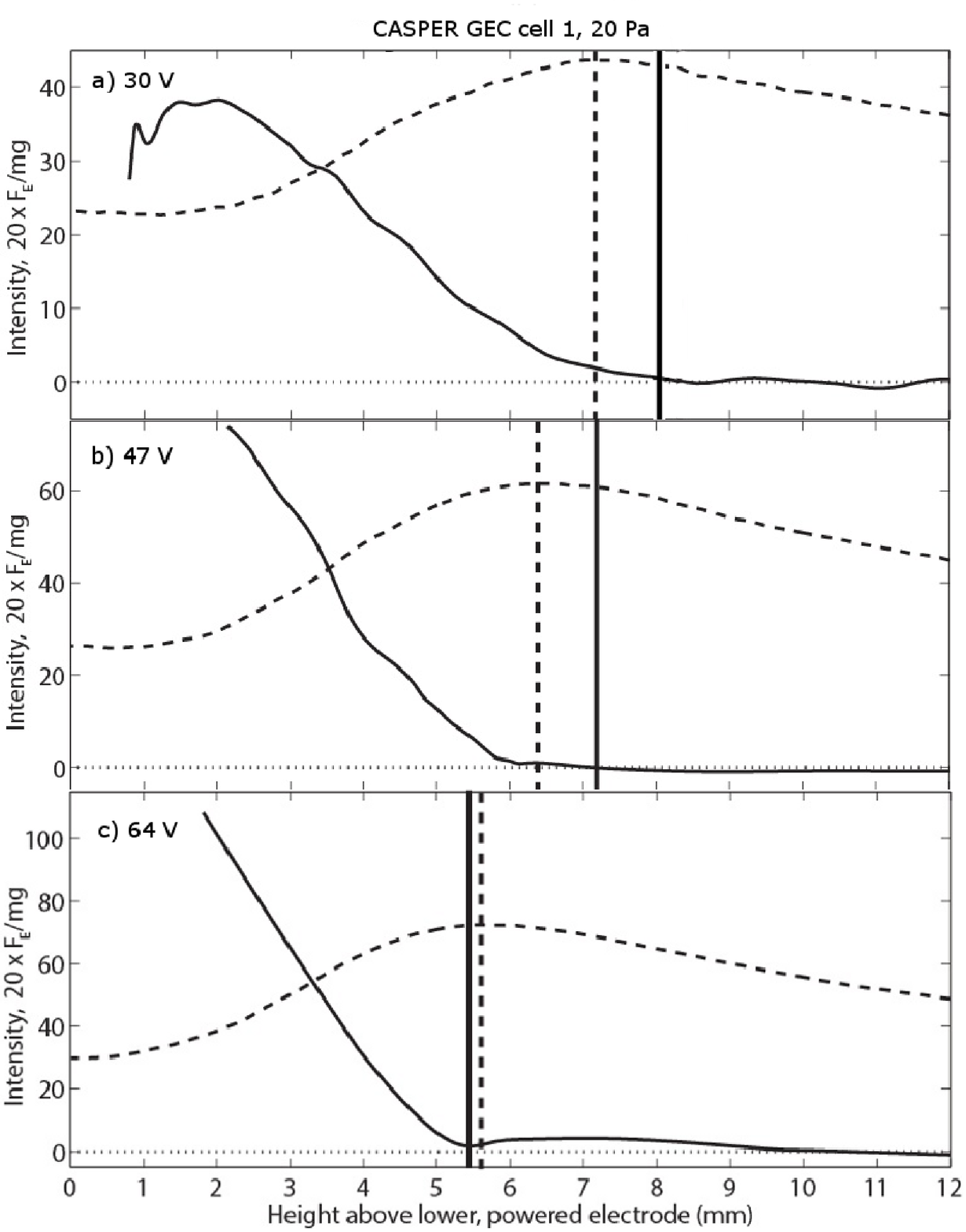}
\caption{The electric force (divided by the gravitational force and multiplied by 20) is indicated by the solid line, and the emission profile (on an 8-bit scale) is indicated by the dashed line for plasma at 20 Pa and for $V_{RF}$ of a) 30, V b) 47 V, and c) 64 V. The solid vertical line indicates breaking of quasi-neutrality, while the dashed vertical line indicates the maximum intensity in the glow profile.}
\label{fig:forces_glow_combo3}
\end{figure}

It has been shown that electric force profiles can be used to locate the sheath edge \cite{Douglass2012}; however, optical emission from a plasma (when no dust is present) is also commonly used to determine the sheath edge. The dominant transition contributing to the optical emission in argon plasma at the low pressures and powers used in the following experiment is the atomic 5p-4s transition. This results in emitted light with a wavelength of 450 nm and gives argon plasma its characteristic indigo color. For comparison of these two methods used to locate the sheath edge, the optical emission obtained from a plasma without dust and the electric force acting on dust (normalized by $20/m_dg$) as determined from single-particle trajectories are plotted in Figure \ref{fig:forces_glow_combo3}. The optical emission profile was obtained by optical emission spectroscopy in which the optical light emitted from the plasma was collected by a CCD camera without the use of filters or other spectroscopic tools. The intensity of each pixel (on an 8-bit scale) was then averaged across each horizontal row in order to obtain the emission intensity as a function of the height above the lower electrode. Figure \ref{fig:forces_glow_combo3}a shows both curves obtained at a system pressure of 20 Pa and a driving potential amplitude of $V_{RF}=30$ V. The solid vertical line indicates the point where quasi-neutrality breaks, as determined by the growth of the electric force acting on the dust, which corresponds to the sheath edge \cite{Douglass2012}. The dashed vertical line indicates the point of maximum glow intensity. Figure \ref{fig:forces_glow_combo3}b and Figure \ref{fig:forces_glow_combo3}c show similar plots for $V_{RF}=47$ V and $V_{RF}=64$ V, respectively. 

The uncertainty in the sheath edge location determined from the force profiles is roughly an electron Debye length \cite{Douglass2012}, ${\lambda}_e = \sqrt{{\epsilon}_0k_bT_e/e^2n_e} \approx 0.5$ mm. Therefore, within this uncertainty, the maximum emission coincides with the sheath edge. With increasing power (increasing $V_{RF}$), the sheath edge and maximum intensity shift toward the electrode and the coincidence seems to improve.

\subsection{Glow and the onset of dust density waves}\label{sec:glowandddw}
\begin{figure}[!htbp]
\centering
\includegraphics[width=3.5in]{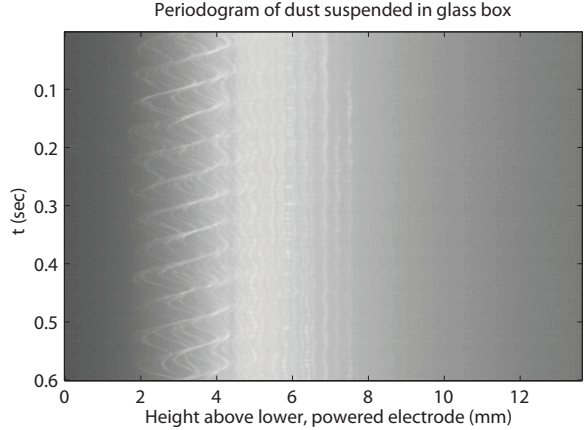}
\caption{A periodogram of dust particles suspended in a glass box, perturbed by DDWs. Time increases from top to bottom. White curves show dust particle trajectories. The horizontal axis corresponds to the vertical direction in the experiment and shows the distance from the electrode.}
\label{fig:periodogram}
\end{figure}

In a second experiment, dust clouds were suspended in a glass box placed on top of the powered, lower electrode in order to achieve dust aggregation in the laboratory. Self-sustained dust density waves (DDWs) occur when the neutral gas pressure within the system is suddenly decreased, providing the relative velocities necessary for the charged dust grains to collide \cite{Matthews2011}. Figure \ref{fig:periodogram} shows a periodogram of dust particles moving in such a cloud when DDWs were present at a pressure of 40 Pa and a power of roughly 4 Watts. 

A wave-pattern is visible with growing amplitude toward the electrode, while farther away from the electrode only very small amplitude oscillatory motion is visible, if at all. The period of the wave motion near the electrode is about 0.075 seconds, which results in a wave frequency of approximately 13 Hz. This frequency is consistent with observed wave frequencies in Q-machines and RF discharges \cite{Barkan1995,Chu1994}. 

\begin{figure}[!htbp]
\centering
\includegraphics[width=3.2in]{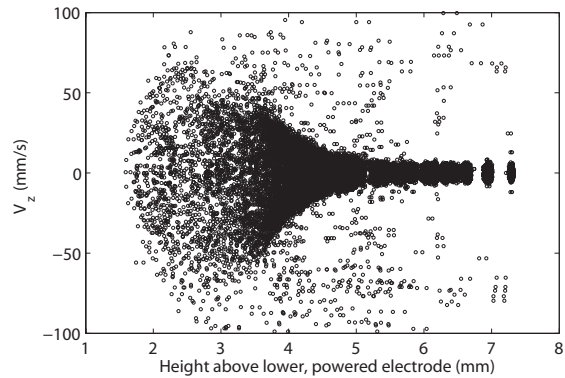}
\caption{Vertical dust particle velocities obtained by tracking particles in the high-speed camera images. Only trajectories that extended over at least a period were included.}
\label{fig:velocities}
\end{figure}

Although single particles cannot be individually resolved in Figure \ref{fig:periodogram}, individual frames were used to locate single particles and track their position. This was possible due to the high frame rate used. From the position data, the vertical velocities as a function of height ($v_z(z)$) were calculated and are shown in Figure \ref{fig:velocities}. Only trajectories that extended over at least one period were included. 

\begin{figure}[!htbp]
\centering
\includegraphics[width=3.5in]{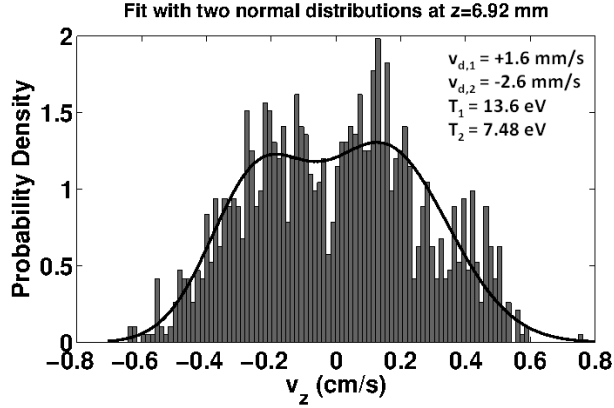}
\caption{The vertical velocity distribution at a distance of 6.92 mm above the electrode. The population consists of particles moving up and down with average velocity $v_{d,1}$, $v_{d,2}$, respectively, having temperatures $T_1$, $T_2$.}
\label{fig:vdist-675}
\end{figure}

\begin{figure}[!htbp]
\centering
\includegraphics[width=3.5in]{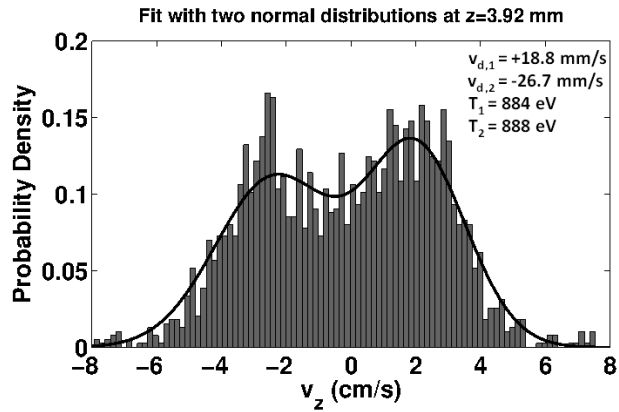}
\caption{The vertical velocity distribution at a distance of 3.92 mm above the electrode. The population consists of particles moving up and down with velocity $v_{d,1}$, $v_{d,2}$, respectively, having temperatures $T_1$, $T_2$.}
\label{fig:vdist-975}
\end{figure}

It is evident from Figure \ref{fig:velocities} that the range of dust particle velocities varies significantly with height above the lower electrode. The sudden increase in the width of the vertical particle velocity distribution (ignoring the outliers outside the main envelope) coincides with the sudden growth of the wave amplitude seen in Figure \ref{fig:periodogram}. This occurs between 4 and 6 mm above the electrode surface. In order to illustrate this variation, the dust velocity distribution at a point above this boundary (z = 6.92 mm) and a point below this boundary (z = 3.92 mm) are shown in Figures \ref{fig:vdist-675} and \ref{fig:vdist-975}, respectively. It is important to realize that the distributions shown in these figures do not represent a snapshot of particle velocities at one given moment in time, but rather they consist of the velocities of many particles at \emph{different times} at \emph{one fixed position}. Therefore, both particles moving up (away from the electrode) and down (toward the electrode) are observed, but at different times. It is the relative vertical motion of particles at different heights that causes the dust compression and rarefaction observed as the wavefronts.

The dust velocity distribution of all particles at a specific height (for all times) cannot be fitted by a single distribution, but is statistically well fitted by a bimodal normal distribution, as shown in the figures. The mean of each normal distribution gives the average velocity of a population of particles (positive values correspond to upward motion, negative to downward). The bimodal fit indicates that the average velocity of the particles moving upward is in fact different from the average velocity of the particles moving downward. The width of the distributions is an indication of the random thermal motion around this average dust velocity and is determined from the standard deviation provided by the fits through ${\sigma}^2 = k_bT_d/m_d$.

\begin{figure}[!htbp]
\centering
\includegraphics[width=3.5in]{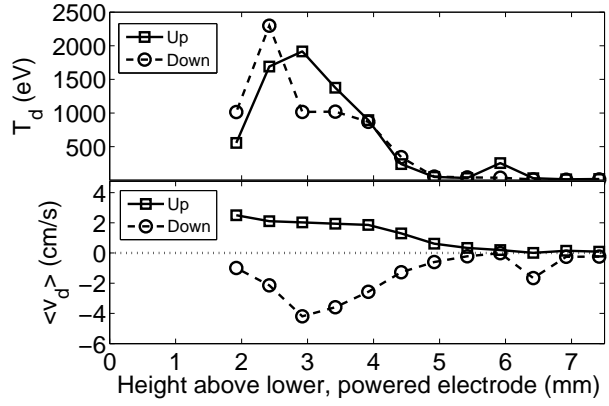}
\caption{Upper panel: thermal energies in eV of the two populations (moving up indicated by the solid line, moving down by the dashed line). Lower panel: average velocities of the particles (moving up indicated by the solid line, moving down by the dashed line).}
\label{fig:temps}
\end{figure}

Vertical velocity distribution graphs, such as those shown in Figures \ref{fig:vdist-675} and \ref{fig:vdist-975}, were created across the entire height of the dust cloud at 0.5 mm increments. From these graphs, the dust temperature and average velocity of the two populations (both upward and downward moving dust) at each height were obtained. The dust temperature and average velocity are plotted as a function of height in Figure \ref{fig:temps}. 

Roughly 5 mm above the lower electrode, both populations show increasing average dust velocities and increasing thermal energy. The magnitude of the downward average dust velocity is slightly larger than the upward average dust velocity for most heights, while the thermal energies are similar for both populations and reaches a maximum value of approximately 2000 eV. The decrease in temperature and velocity profiles near the lowest data point may be due to the limited number of trajectories observed within the time frame and the limited amount of data available.

\begin{figure}[!htbp]
\centering
\includegraphics[width=3.5in]{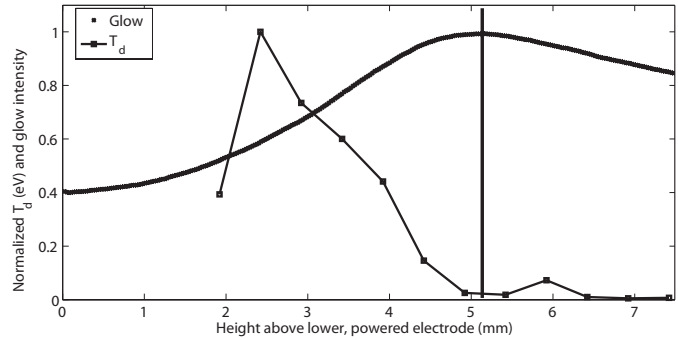}
\caption{The average dust temperature of the two populations, normalized to 2000 eV, and the normalized glow intensity. A sudden rise in the dust temperature occurs at the same height where the glow reaches the highest intensity, indicated by the vertical line.}
\label{fig:dusttemp}
\end{figure}

Finally, the average dust temperature (averaged over the upward and downward populations) is plotted together with the intensity of the plasma glow (obtained without dust) in Figure \ref{fig:dusttemp}. It is clear the sudden increase in temperature of the particles in the DDWs occurs at the height where the emission reaches its maximum value, which should correspond to the sheath edge, as determined in the previous experiment.

\section{Conclusions and discussion}
The time-averaged glow is a valuable additional source of information for determining physical processes occurring within boundary layers of plasma (i.e. the sheath). We can summarize our main conclusions as follows:

\begin{enumerate}
\item{The maximum intensity of the time-averaged optical emission lies at, or within an electron Debye length of the point where quasi-neutrality is broken, i.e. the sheath edge.}
\item{The dust temperatures and average velocities of particles in DDWs show a sudden increase at the point of maximum glow intensity, hence at the sheath edge.}
\end{enumerate}

The sudden increase in dust temperature at the point where the wave amplitude and average particle velocities also suddenly increase is likely an indicator of the instability responsible for the waves. There are multiple mechanisms that have been advanced as probable causes for the onset and growth of DDWs (or dust acoustic waves, DAWs): the ion-streaming instability \cite{Rosenberg1996}, a dust charge gradient (in the case of sufficient dust density and charge-dependent external forces) \cite{Shukla2000}, and in one case the presence of a small constant electric field across the dust cloud, even without a charge gradient, but only at high dust densities \cite{Fortov2003}.

Local breaking of the linear stability criterion for the ion streaming instability, caused by ions accelerated downwards in the sheath interacting with the dust particles, requires the local ion drift velocity to be significantly larger than the ion thermal velocity ($u_{+}(z)\gg v_T$) \cite{Rosenberg1996}. Putting this in terms of the Bohm velocity, $v_B=\sqrt{T_e/m_{+}}$, we find $u_{+}(z)/v_B\gg v_T/v_B=\sqrt{T_{+}/T_e}\approx$0.1. Hence, $u_{+}(z)\gg0.1v_B$. This implies that the local ion drift velocity should approximately equal the Bohm velocity in order for DDWs to appear, if caused by the linear ion streaming instability.

Recent simulations in geometries similar to our experiments have shown that the Bohm point (the height where $u_{+}=v_B$) can lie well \emph{below} the point where quasi-neutrality is broken, hence where the sheath edge is located \cite{Douglass2012, Loizu2011}. These simulations also showed that the local electric field at this point is on the order of E $\approx$ 5 V/cm. The mean free path for an argon ion is given by $l_{mfp}=(n_g\sigma)^{-1}$. With a gas density of $n_g\approx 5$x$10^{21}$m$^{-3}$ and the Ar$^+$-Ar momentum transfer cross section given by $\sigma \approx 10^{-18}$m$^{2}$ \cite{Phelps1990}, the argon ion mean free path is found to be $l_{mfp}\approx 0.2$ mm. Assuming all momentum is lost between collisions, an argon ion will obtain a kinetic energy between collisions by being accelerated in the local electric field over a mean free path; $\frac{1}{2}m_{+}u_{+}^2=qE\Delta z = eEl_{mfp}\approx$ 0.1 eV. The Bohm energy equals $T_e/2 \approx$ 2 eV, therefore, ions accelerated in the small electric field near the sheath edge (defined by the point where quasi-neutrality is broken) are unlikely to attain Bohm velocity. This implies that the linear stability criterion cannot be broken in this region and is therefore not responsible for the instability observed.

On the other hand, the computed electric field mentioned above is similar to the electric field of 3 V/cm required for the onset of DAWs, as reported for dust clouds with very high dust density \cite{Fortov2003}. The onset of waves observed in the glass box used in this experiment only occur when a large number of dust particles are present, which is consistent with this observation \cite{Shukla2000}. 



Simultaneously, the same simulations have shown that the dust charge varies strongly within the sheath and the variation in the dust surface potential, ${\Phi}_D$, with vertical position can be as high as 2 V/mm, while the equilibrium potential is roughly -10 V. The large amplitude waves have an amplitude on the order of a millimeter, therefore, $\left(\Delta {\Phi}_D/{\Phi}_D\right) \sim 0.2$ during an oscillation. This implies that the change in the charge on a dust particle during an oscillation is on the order of $\pm$ 20\%, which is rather large. Large charge variations within the cloud can cause further growth of any unstable modes, which might help explain the onset of the instabilities.


In short, the observed coincidence of the onset of the DDWs with the location of the sheath edge, together with the numerical simulation results, indicate that the ion streaming instability is not the driving factor behind the DDWs. On the other hand, the dust charge variation due to the oscillatory motion of the dust, together with the local electric field near the sheath edge and the high dust density could very well be the determining factors for the onset and subsequent growth of the waves, at least for the geometry considered here. However, the simple techniques presented here can easily be used to determine if these conclusions are valid in other systems as well.

\section*{Acknowledgment}

The authors would like to thank Jorge Carmona-Reyes, Jimmy Schmoke, and Mike Cook for their support. We would also like to thank the referees for their valuable suggestions that served to improve the manuscript.

\end{document}